\documentclass{aa}
\usepackage{graphicx}
\usepackage{txfonts}

\begin{document}

\title{Infrared and visual lunar occultations measurements of
stellar diameters and new binary stars detections at the
Calar Alto 1.5\,m telescope}

\author{O. Fors\inst{1}\fnmsep\inst{2}
\and A. Richichi\inst{3}
\and J. N\'u\~nez\inst{1}\fnmsep\inst{2}
\and A. Prades\inst{4}
}

\institute{
Departament d'Astronomia i Meteorologia, Universitat de  Barcelona,
Av. Diagonal 647, 08028 Barcelona, Spain
\and
Observatori Fabra, Cam\'{\i} de l'Observatori s/n, 08035 Barcelona, Spain
\and
European Southern Observatory,
Karl-Schwarzschild-Str. 2, D-85748 Garching bei M\"unchen, Germany
\and
Escola Universitaria Polit\`ecnica de Barcelona,
Universitat Polit\`ecnica de Catalunya, E-08028, Barcelona, Spain
}
\offprints{O. Fors, \email{ofors@am.ub.es}}
\date{Received / Accepted }
\titlerunning{Stellar diameters measurements and new binary stars detection}

\abstract{
We present a program of routine lunar occultations, at optical and near-IR
wavelengths, recently started at the 1.5\,m Spanish telescope at the
Calar Alto Observatory. Both a CCD and an infrared array detector are
used. The program is aimed mainly at the detection and investigation of
binary systems, although results in other areas of stellar research are
also anticipated.
Occultations are reported for a total of 40 stars. Among these, {\object
{SAO 164567}}, {\object {SAO 78258}} and {\object {AG+24~788}} have been  
discovered to be binaries, with projected separations as small as
$0\farcs006$. Furthermore, binarity is suspected in the case of {\object
{SAO 78119}} and {\object {SAO 79251}}. Additionally, the angular diameter
of the late-type giant {\object {30~Psc}} and of the infrared star
{\object {V349 Gem}} have been accurately measured, this latter for the
first time.
We finally evaluate the instrumentation performance in terms of limiting
magnitude and angular resolution, and discuss applications to larger telescopes.
}
\maketitle

\section{Introduction}
The technique of lunar occultations (LO) has established itself  as a very
simple and powerful means to achieve high angular resolution. It consists
essentially in performing high-speed photometry of the light curve produced
when a background star is occulted by the limb of the Moon. Angular resolution
information on the occulted source at the scale of one milliarcsecond (mas) can
be retrieved from the analysis of the diffraction fringes, either through
model-dependent or model-independent approaches (Richichi \cite{richichi89}).
Since the occultation is a diffraction phenomenon produced at the lunar limb
rather than in the telescope, these observations are characterized by several
properties which are significantly different from the ones typically
encountered with other techniques for high angular resolution such as adaptive
optics, speckle or long-baseline interferometry. For example, the limiting
angular resolution  is not set by the size of the telescope nor by the
wavelength of observation, at least to a first approximation. Also, it is not
directly influenced by the quality of the seeing.

While these properties make the LO technique very attractive, a number of
shortcomings limit its application considerably. First and foremost,
occultations can be observed only for sources which lie on the apparent orbit
of the Moon. This restricts the candidates to a narrow belt around the Zodiac,
covering approximately 10\% of the celestial sphere.  Second, LO are fixed-time
events, which need careful planning and the successful combination of technical
and meteorological readiness. Finally, each LO event only provides a
one-dimensional scan of the source, along a direction which is determined by
the lunar motion and the source position. Depending on the conditions, a given
source can be occulted only once or several times over a period of a few months
or very few years. In this case, some limited two-dimensional information can
be obtained.

The combination of these advantages and limitations, makes the LO technique
particularly appealing especially for small and medium-sized telescopes, where
a routine program of observations can be established. The availability of
relatively cheap detectors for visual and near-infrared (NIR) fast photometry
is bringing this method within the budget of most observatories. In this
respect, we note the benefits of commercial, relatively cheap CCDs developed
for amateur astronomers on one side, and previous-generation NIR arrays of
small format on the other side.  In their respective wavelenghts ranges, both
kind of detectors offer a quality sufficient for the purpose of LO
observations, where the noise is essentially set by the lunar background.

These considerations have motivated us to start a LO program at the 1.5\,m
Spanish telescope of the Calar Alto Observatory. This program is independent of
previous similar efforts at the same observatory, in which some of us were
involved. Unfortunately, those efforts are presently stalled, given the
difficulty of allocating time to the required instrumentation. The present
program is carried out at a different telescope and with different
instrumentation. In addition to the allocation of sufficient observing time,
the program is finding its success thanks to the availability of two
instruments for which specific adaptations allow us a rapid sampling of the
lightcurves, as decribed in Sect.~\ref{data}. In Sect.~\ref{results} we
describe the first results obtained on a number of sources, including new
binary detections and stellar angular diameters. In Sect.~\ref{performance}, we
evaluate the performance of the two instruments employed, and extrapolate their
possible application to larger telescopes.

\section{Observations and data reduction}\label{data}
Visual and NIR observations were carried out with the 1.5\,m telescope of the
Observatorio Astronomico Nacional in Calar Alto (Spain) during October 2001 and
February 2002, respectively. For the visual occultations we employed a Texas
Instruments TC-211 CCD, which is assembled as the guiding chip of the SBIG-ST8
camera. This was operated in fast drift-scanning mode as described by Fors et
al. (\cite{fors01}), allowing us to sample the occultation lightcurves at
millisecond rates. For the NIR observations, we made use of the MAGIC camera
(Herbst et al. \cite{herbst93}), operated in fast subarray mode with a window
size of 8x8 pixels. This size was found to provide a good compromise between a
sufficiently fast sampling rate (typically 8.5\,ms with an integration time of
3\,ms) and a robust estimation of the background level around the stellar
image.

Table~\ref{table_list} lists the parameters of the observed occultations,
following closely the format established in Richichi et al. (\cite{bina6}), and
previous papers of that series. Columns (1) through (3) list the source
identification, the date of the event and the configuration used.  The code CA
refers to observations with the CCD, and the code CB to observations with
MAGIC. Broad band R ($641\pm58$nm) and K ($2.2\pm0.4\mu$m) filters
were used, respectively. Column (4) lists the field of view either set by the
diaphragm aperture or by the array subwindow. Columns (5) and (6) list the
sampling time of the lightcurves, and the integration time for each data point.
Columns (7) through (9) list the total magnitude of the star in the V, R and K
filters.  The V magnitudes are taken from the literature. In principle,
photometric information could be extracted directly from the LO data. However,
we lacked  an accurate instrument calibration, and we wished to avoid possible
systematic biases: this is the case of the CCD drift-scanning technique,  since
the object does not effectively fit within the single column which is readout
at each integration time.  Instead, we have collected the R and K magnitudes
from the USNO-B1.0 and 2MASS catalogues, respectively. Note that many of the CB
sources have K magnitudes above the saturation limit of 2MASS. Although this
has been accounted for in the catalogue, systematic errors might still be
present and possible examples are discussed in Sect~\ref{SNR}. In column (10)
we report the spectral types, again extracted when available from the
literature; in the case of multiple determinations, the most frequent or most
recent was used. Finally, column (11) lists the distances based on Hipparcos
parallaxes, when available. Those values affected by a large uncertainty
(>10\%) have been omitted. 

\begin{table*}
\caption{List of the occultation events and of the circumstances of their observation\label{table_list}}
\begin{tabular}{lccrrrrrllcl}
\hline 
\multicolumn{1}{c}{(1)}&
\multicolumn{1}{c}{(2)}&
\multicolumn{1}{c}{(3)}&
\multicolumn{1}{c}{(4)}&
\multicolumn{1}{c}{(5)}&
\multicolumn{1}{c}{(6)}&
\multicolumn{1}{c}{(7)}&
\multicolumn{1}{c}{(8)}&
\multicolumn{1}{c}{(9)}&
\multicolumn{1}{c}{(10)}&
\multicolumn{1}{c}{(11)}\\
\multicolumn{1}{c}{Source}&
\multicolumn{1}{c}{Date}&
\multicolumn{1}{c}{Tel.+}&
\multicolumn{1}{c}{D}&
\multicolumn{1}{c}{$\Delta $t }&
\multicolumn{1}{c}{$\tau $ }&
\multicolumn{1}{c}{V}&
\multicolumn{1}{c}{R}&
\multicolumn{1}{c}{K}&
\multicolumn{1}{c}{Sp.}&
\multicolumn{1}{c}{Dist.}\\
&
\multicolumn{1}{c}{UT}&
\multicolumn{1}{c}{detector}&
\multicolumn{1}{c}{$\arcsec $}&
\multicolumn{1}{c}{ms}&
\multicolumn{1}{c}{ms}&
\multicolumn{1}{c}{mag}&
\multicolumn{1}{c}{mag}&
\multicolumn{1}{c}{mag}&
\multicolumn{1}{c}{}&
 pc&
\\
\hline 
\object{SAO 187645} & 22-10-01 & CA & 8 & 2.2 & 2.2 & 8.3 & 7.4 && K1/K2III &       \\
\object{SAO 187660} & 22-10-01 & CA & 11& 2.2 & 2.2 & 7.3 & 6.5 && K2III    &   128 \\
\object{SAO 189746} & 24-10-01 & CA & 6 & 1.8 & 1.8 & 8.6 & 8.0 && G8III    &       \\
\object{SAO 189774} & 24-10-01 & CA & 8 & 5.0 & 5.0 & 9.5 & 8.8 && K1/K2III &       \\
\object{SAO 164553} & 25-10-01 & CA & 7 & 5.0 & 5.0 & 8.5 & 8.3 && F0III/IV &       \\
\object{SAO 164567} & 25-10-01 & CA & 8 & 1.8 & 1.8 & 7.4 & 6.5 && K5III    &       \\
\object{SAO 165121} & 26-10-01 & CA & 7 & 3.0 & 3.0 & 9.2 & 8.5 && K1/K2III &       \\
\object{SAO 165128} & 26-10-01 & CA & 7 & 6.0 & 6.0 & 9.5 & 9.0 && G2/G3V   &       \\
\object{SAO 165136} & 26-10-01 & CA & 6 & 2.0 & 2.0 & 7.8 & 7.0 && K0III    &   230 \\
\object{SAO 165578} & 27-10-01 & CA & 6 & 2.1 & 2.1 & 6.1 & 5.3 && K5III    &   256 \\
\object{SAO 147033} & 28-10-01 & CA & 7 & 1.6 & 1.6 & 7.7 & 6.9 && K0       &   238 \\
\object{SAO 147032} & 28-10-01 & CA & 7 & 1.5 & 1.5 & 7.8 & 7.5 && F5       &   221 \\
\object{30 Psc}     & 28-10-01 & CA & 6 & 1.5 & 1.5 & 4.4 & 3.6 && M3III    &   127 \\
\object{SAO 78001}  & 22-02-02 & CB & 7 & 8.5 & 3.0 & 9.1 && 7.7 &  F0      &       \\
\object{SAO 78119}  & 22-02-02 & CB & 7 & 8.7 & 3.0 & 8.1 && 4.9 &  K0      &       \\
\object{SAO 78122}  & 22-02-02 & CB & 7 & 8.4 & 3.0 & 7.9 && 5.7 &  G5      &   217 \\
\object{SAO 78168}  & 22-02-02 & CB & 7 & 8.4 & 3.0 & 6.1 && 3.9 &  G8III   &   134 \\
\object{SAO 78176}  & 22-02-02 & CB & 7 & 8.4 & 3.0 & 6.3 && 4.9 &  B3Ib    &       \\
\object{SAO 78192}  & 23-02-02 & CB & 7 & 8.4 & 3.0 & 8.4 && 3.6 &  M...    &       \\
\object{SAO 78197}  & 23-02-02 & CB & 7 & 8.6 & 3.0 & 8.2 && 5.3 &  K0      &       \\
\object{DO 12097}   & 23-02-02 & CB & 7 & 8.4 & 3.0 & 9.3 && 5.3 &          &       \\
\object{SAO 78210}  & 23-02-02 & CB & 7 & 8.5 & 3.0 & 6.6 && 4.5 &  G5      &   242 \\
\object{V349 Gem}   & 23-02-02 & CB & 7 & 8.3 & 3.0 &12.2 && 4.1 &          &       \\
\object{SAO 78258}  & 23-02-02 & CB & 7 & 8.5 & 3.0 & 8.2 && 6.9 &  G0      &   198 \\
\object{SAO 78272}  & 23-02-02 & CB & 7 & 8.5 & 3.0 & 7.3 && 5.0 &  K0      &       \\
\object{SAO 79133}  & 23-02-02 & CB & 7 & 8.5 & 3.0 & 7.9 && 6.8 &  F5      &    72 \\
\object{AG+24 788}  & 23-02-02 & CB & 7 & 8.4 & 3.0 & 10.3&& 6.4 &  K0      &       \\
\object{SAO 79162}  & 23-02-02 & CB & 7 & 8.5 & 3.0 & 5.9 && 4.8 &  F5III-IV&   107 \\
\object{SAO 79176}  & 23-02-02 & CB & 7 & 8.5 & 3.0 & 9.2 && 7.6 &  G5      &       \\
\object{SAO 79194}  & 23-02-02 & CB & 7 & 8.5 & 3.0 & 8.7 && 7.5 &  F5      &       \\
\object{SAO 79214}  & 23-02-02 & CB & 7 & 8.5 & 3.0 & 7.9 && 5.6 &  G5      &   236 \\
\object{SAO 79236}  & 23-02-02 & CB & 7 & 8.5 & 3.0 & 8.1 && 7.0 &  F8      &    40 \\
\object{SAO 79251}  & 23-02-02 & CB & 7 & 8.5 & 3.0 & 8.7 && 6.3 &  K0      &       \\
\object{SAO 79257}  & 23-02-02 & CB & 7 & 8.5 & 3.0 & 8.4 && 7.4 &  F5      &   167 \\
\object{AG+24 824}  & 23-02-02 & CB & 7 & 8.5 & 3.0 & 10.2&& 7.9 &  G0      &       \\
\object{AG+23 808}  & 24-02-02 & CB & 7 & 8.4 & 3.0 & 10.2&& 7.8 &  K0      &       \\
\object{SAO 79302}  & 24-02-02 & CB & 7 & 8.5 & 3.0 & 8.3 && 7.9 &  A2      &   280 \\
\object{SAO 79325}  & 24-02-02 & CB & 7 & 8.5 & 3.0 & 9.5 && 7.0 &  K2      &       \\
\object{SAO 79365}  & 24-02-02 & CB & 7 & 8.5 & 3.0 & 9.3 && 6.2 &  K7      &       \\
\object{IRAS 07231+2349} & 24-02-02 & CB & 7 & 8.4 & 3.0 &&&4.0 &          &        \\
\hline                                                                      
\end{tabular}                                                               
\\                                                                          
\end{table*}

The data were analyzed with the same methods used by Richichi et al.
(\cite{bina6}, \cite{diam6}), and other papers in those series. In particular,
a model-dependent least squares method was employed, which uses as free
parameters the angular diameter for single stars, and additionally for binary
stars the  angular diameter of the companion, the projected separation and the
brightness ratio. Other free parameters include the rate of the event, the
intensity of the background and its time drift. Spurious frequencies due to
pick-up of mains power and other effects have been occasionally noticed, and
have been digitally filtered. Our data analysis can also include a fit and
removal of relatively slow, random fluctuations of the background (due to thin
cirrus and lunar halo) and of the stellar intensity (due to image motion and
scintillation), by means of modelling through Legendre polynomials as described
in the above mentioned papers.

We also used a model-independent method (CAL, Richichi \cite{richichi89}), which is
particularly suited in detecting companions at very small separations and the
presence of extended circumstellar emission.

\section{Results}\label{results}
The stars for which a positive result could be obtained are listed in
Table~\ref{table_results}.  In the table, the format follows the same style as
in Richichi et al. (\cite{bina6}) and other papers of that series. A detailed
discussion of the listed quantities is given in Richichi et al. (\cite{bina2}).
In summary, the columns list the absolute value of the fitted linear rate of
the event V in $m~ms^{-1}$, its deviation from the predicted rate V$_{\rm{t}}$ as
computed by us, the local lunar limb slope $\psi $, the true position and
contact angles, the signal--to--noise ratio (SNR). For binary detections, the
projected separation and the brightness ratio are given, while for the single
stars the angular diameter $\phi_{\rm UD}$ is reported, under the assumption of
a uniform stellar disc. Only in the case of  \object{SAO 164567}, the
predicted contact angle was sufficiently close to zero that even a small
difference of 3\% between predicted and measured rate results in an imaginary
value of the limb slope. In this case, we list the predicted, rather than
measured, PA and CA values.

\begin{table*}
\caption{Summary of results\label{table_results}}
\begin{tabular}{lrrrrrrrrr}
\hline 
\hline 
\multicolumn{1}{c}{(1)}&
\multicolumn{1}{c}{(2)}&
\multicolumn{1}{c}{(3)}&
\multicolumn{1}{c}{(4)}&
\multicolumn{1}{c}{(5)}&
\multicolumn{1}{c}{(6)}&
\multicolumn{1}{c}{(7)}&
\multicolumn{1}{c}{(8)}&
\multicolumn{1}{c}{(9)}&
\multicolumn{1}{c}{(10)}\\
\multicolumn{1}{c}{Source}&
\multicolumn{1}{c}{$|$V$|$}&
\multicolumn{1}{c}{V/V$_{\rm{t}}$--1}&
\multicolumn{1}{c}{$\psi $}&
\multicolumn{1}{c}{PA}&
\multicolumn{1}{c}{CA}&
\multicolumn{1}{c}{SNR}&
\multicolumn{1}{c}{Sep. (mas)}&
\multicolumn{1}{c}{Br. Ratio}&
\multicolumn{1}{c}{$\phi_{\rm UD}$ (mas)}\\
\hline 
\object{SAO 164567} & 0.6443 & 3\% &      &(74)&(11) & 14.3 &  2.0 $\pm$ 0.1 &  1.7 $\pm$ 0.1     &  \\
\object{30 Psc} & 0.2473 & $-44$\% & 20  & 122 & 69  & 46.1 &                &                    &$6.78\pm0.07$  \\
\hline 
\object{SAO 78119} & 0.5387 &$-3$\% & $2$& 129 &  41 & 52.7 & 13.1 $\pm$ 1.1 & 34.2 $\pm$ 2.5     &  \\
\object{V349 Gem} & 0.9462 & $-2$\% &  8 & 106 & 11 & 65.9 &                &                    & $5.10 \pm 0.08$ \\
\object{SAO 78258} & 0.6307 & 2\%   &  1 &  45 &$-50$& 9.4 & 47.3 $\pm$ 1.5 & 8.6 $\pm$ 0.7     &  \\
\object{AG+24 788} & 0.6910 & 3\% & 6 & 75 &$-13$& 16.9 &28.8 $\pm$ 0.7 &  4.9 $\pm$ 0.2     &  \\
\object{SAO 79251} & 0.7215 &$-1$\% & $-1$&  85 &$-15$& 20.2 &26.9 $\pm$ 1.1 & 17.6 $\pm$ 1.5     &  \\
\hline 
\hline 
\end{tabular}
\par{}
\end{table*}

\subsection{\object{SAO~164567}}
This star was included in observations both by the Hipparcos satellite, and by
radial velocity measurements (Moore \& Paddock \cite{moore50}, Duflot et al.
\cite{duflot95}), but it was never reported as binary.

\subsection{\object{30~Psc}}
This long period variable is classified  as oxygen-rich AGB MIII giant without
dust emission (Sloan \& Price \cite{sloan98}). It has been catalogued by
Hipparcos as suspected non--single. One speckle observation with a limiting
resolution of $0\farcs054$ was inconclusive in this respect (Mason et al.
\cite{mason99}).  Two interferometric observations in the K band by Dyck et al.
(\cite{dyck98}) led to an angular diameter of $7.2\pm0.5$\,mas. This is in
agreement, within the error bars, with our estimation given in
Table~\ref{table_results}. Further observations would be useful to assess
whether there is a measurable dependence of the angular diameter with
wavelength. Neither the interferometric measurements nor our LO seem to
indicate evidence of binarity.

\subsection{\object{SAO~78119}, \object{SAO~78258}, \object{AG+24~788},
\object{SAO~79251}} 
No previous indications of binarity for these stars are reported in the
literature. {\object{SAO~78258} was listed in the Hipparcos catalogue as single
star.

\subsection{\object{V349~Gem}}
No previous high-angular resolution measurements are listed for this carbon
star in the literature. Epchtein et al.(\cite{epchtein87}) have classified it
as having a temperature below 2500\,K, and with a circumstellar dust shell of
400-1500\,K.  Our value for the angular diameter of this star seems to  be
consistent with the general properties of this star. However the spectral
energy distribution of this source is poorly known, also in consideration of
its variability, and it is not possible to constrain  significantly the
effective temperature. Further photometric monitoring is desirable.

\subsection{Stars with negative binary detection} Among the stars for which we
did not detect any binarity, a few are known to have companions.  After a close
examination of the characteristics of the stars and a comparison with the
circumstances and achieved performance of our observations, we conclude that
there are no significant discrepancies (see Table~\ref{table_negative} for
a brief explanation of each non-detection). In the following, we provide
a discussion of the individial cases which deserve special attention.

\object{SAO 78168} was reported to have duplicity discovered by visual
occultation (Zhitetski \cite{zhitetski77}). However, such observation was
catalogued as doubtful in XZ80 catalog (Dunham \& Warren \cite{dunham95}) due
to the event was recorded to have a gradual disappearance (Dunham \& Herald
\cite{dunham04}). In addition, no binarity is reported in the Hipparcos catalogue.
Finally, we note that the occultation trace of this star could be recorded at a
very good SNR, permitting us to cover a dynamic range of almost five magnitudes
from the primary.

\object{SAO 79257} is a known subarcsecond binary, reported in the Washington
double star (\object{WDS J07181+2405}) as well as in the Hipparcos (\object{HIP
35344}) catalogues.  However, the entries  are not entirely consistent for what
concerns the position angles. WDS reports PA of $153\degr$ and $132\degr$ for
epochs 1971 and 1991 respectively, with a separation of $0\farcs4$ in both
cases and a magnitude difference of about unity. Hipparcos reports a PA of
$158\degr$ in 1991, with a separation of $0\farcs393$. Given that the scan
angle of our LO event ($74\degr$) was almost orthogonal to the PA reported
above, the differences between Hipparcos and WDS are significant.  We have used
a binary star model to fit our occultation data of \object{SAO 79257}. The
result was that the data are consistent with a binary having a separation of
about 20\,mas and a brightness ratio of about 2\,mag (i.e., at the limit of the
sensitivity permitted by the SNR). The resulting $\chi^2$ was improved by only
4\%  with respect to the case of a single star model, and we cannot claim a
positive detection. Our derived projected separation can be reconciled with the
true separation, if the LO scan direction was about $87\degr$ from the PA of
the binary. This would imply PA close to $161\degr$, which is very close to the
value measured by Hipparcos in 1991. We conclude that our data are not
inconsistent with the presence of the known companion, and indicate that this
latter would have to be significantly redder than the F5 primary. However,
further conclusions are not possible given the uncertainties in the actual PA
of the binary.

\begin{table}
\caption[ ]{Summary of negative detection results}\label{table_negative}
\begin{flushleft}
\begin{tabular}{lrrcl}
\hline
\hline
\multicolumn{1}{c}{(1)} &
\multicolumn{1}{c}{(2)} &
\multicolumn{1}{c}{(3)} &
\multicolumn{1}{c}{(4)} &
\multicolumn{1}{c}{(5)} \\
\multicolumn{1}{c}{Source} &
\multicolumn{1}{c}{$\psi$} &
\multicolumn{1}{c}{PA} &
\multicolumn{1}{c}{SNR} &
\multicolumn{1}{c}{Notes}\\
\hline
\object{SAO~164553}&$-4$& 122 &  2.9 &  Outside field of view\\
\object{SAO~165578/B}& 3  &  22 & 5.9  &  Too faint\\
\object{SAO~165578/C}&&&& Outside field of view\\
\object{SAO~78122} &  6 &  87 & 26.6 &  Large separation\\
\object{SAO~78168} &  0 &  73 & 78.9 &  No details known\\
\object{SAO~78197} & 12 & 110 & 31.8 &  Outside field of view\\
\object{SAO~79257} & 10 &  74 & 6.0  &  Consistent with projection\\
\hline
\hline
\end{tabular}
\end{flushleft}
\end{table}

\section{Performance}\label{performance}
\subsection{Limiting magnitude}\label{SNR}
By plotting the SNR as a function of the magnitude of the occulted star, we can
estimate an empirical relation for the limiting magnitude that can be achieved
by observations with the two instruments, as shown in Figures~\ref{SNR_vs_R}
and \ref{SNR_vs_K}. 

\begin{figure}
\resizebox{\hsize}{!}{\includegraphics{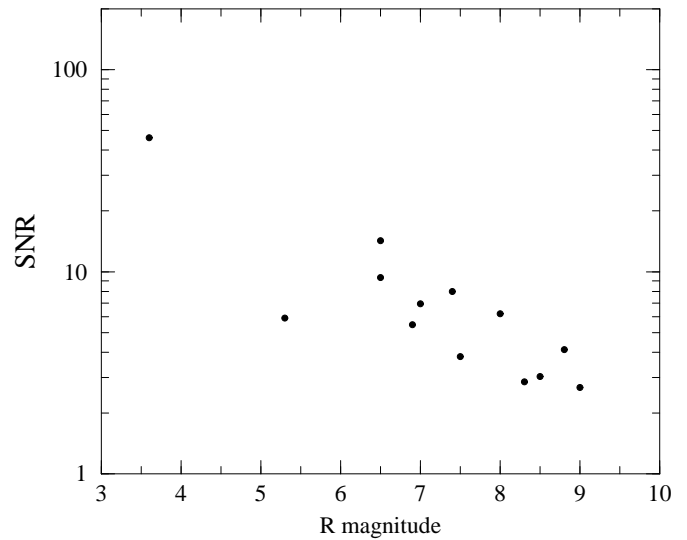}}
\caption{Relation between SNR and R magnitude, for observations with the
SBIG-ST8 instrument.}
\label{SNR_vs_R}
\end{figure}

\begin{figure}
\resizebox{\hsize}{!}{\includegraphics{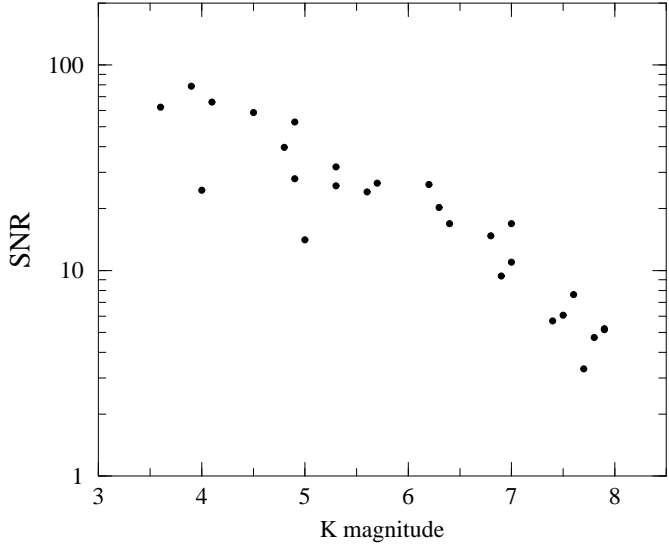}}
\caption{Relation between SNR and K magnitude, for observations with the MAGIC instrument.}
\label{SNR_vs_K}
\end{figure}

It can be noted that in both cases, the data indicate a  clear relationship
between magnitude and SNR. For studies of binary stars, companions with a
brightness ratio close to unity can be detected already when the SNR is
relatively small, in the range 1-3. Figures~\ref{SNR_vs_R} and \ref{SNR_vs_K}
show that LO observations at the 1.5\,m telescope can be used for
investigations of binary systems down to magnitudes R$\approx$9 and
K$\approx$8.5.

It is interesting to compare our result of Fig.~\ref{SNR_vs_K}  with Fig.~3 of
Richichi et al. (\cite{richichi96}), which showed a similar plot for LO data
obtained also with a 1.5\,m telescope (TIRGO) in the K band, but using a fast
photometer. The IR array shows better SNR for the range K$\approx$4-7\,mag,
probably thanks to the ability to reject more background signal and thus reduce
significantly the photon noise in the data. Below K$\approx$7\,mag the
advantage is less clear, due also to the scarcity and scatter of the data
available for a comparison. One possible reason could be that LO events at
TIRGO for such faint sources were recorded under conditions systematically
better than average in terms of background (for example, at low lunar phases).
Our Fig.~\ref{SNR_vs_K}  does not reach magnitudes brighter than K$\approx$3.5,
so no comparison can be drawn at the bright end.

One can also notice a much larger scatter in the figure based on the TIRGO
data. This is probably due to the fact that the TIRGO sample include a much
larger database of LO observations, collected over a wide range of lunar phases
and background conditions, as well as with different  settings of the
photometer. Our data were instead recorded over three consecutive nights, and
with essentially identical settings of subframe size and integration time.

\subsection{Limiting resolution}  
We have also performed an analysis of the limiting angular resolution achieved
in our observations, by both adopting the same definition of resolution
and the same approach described at Richichi et al. (\cite{richichi96}). In
particular, the same code for the automatic estimation of the limiting angular
resolution has been used. This quantity has been computed for 25 unresolved
sources in our sample (this is less than the full size of the sample minus the
resolved sources, because in a few cases the SNR was not sufficient to perform
the computations), and is plotted as a function of SNR in Fig.~\ref{limres}.

\begin{figure}
\resizebox{\hsize}{!}{\includegraphics{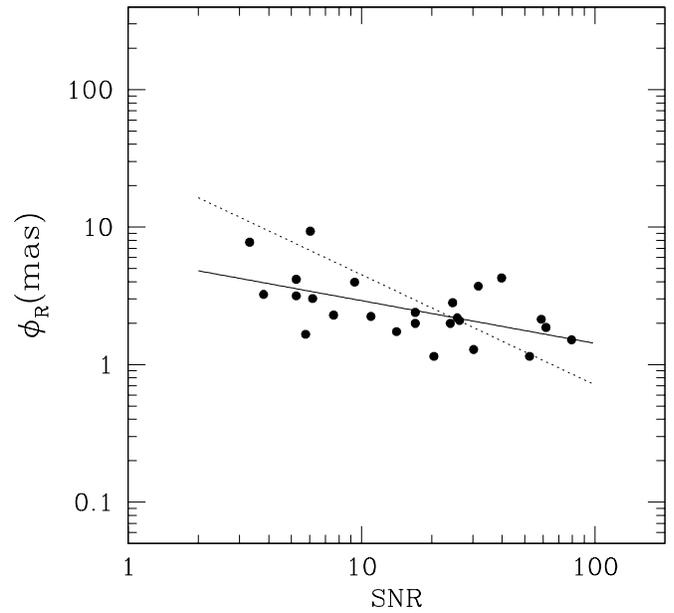}}
\caption{Limiting resolution $\phi_{\rm R}$ for
the unresolved sources in our sample, as a function
of SNR. The solid line represents a linear fit, while
the dotted line is the trend shown in Fig.~5 of 
Richichi et al. (\cite{richichi96}).
}
\label{limres}
\end{figure}

The figure shows, as expected, an improvement in the limiting angular
resolution for increasing SNR. In particular, diameters below 2\,mas are
expected to be resolved for SNR values approaching 100.  However, the slope  
of this function is significantly different from that of LO measurements
obtained with a photometer.  Fig.~\ref{limres} shows the linear approximation
(by least-squares fits) of our data, and the same from Fig.~5 of Richichi et
al. (\cite{richichi96}) for the TIRGO telescope. It can be appreciated how the
data obtained with CCDs and IR arrays, as in the present paper, provide a
better performance in terms of limiting angular resolution in the low SNR
regime. This is due to the better performance  on faint sources, as already
discussed in Sect.~\ref{SNR}. However, the data obtained with the TIRGO
photometer provide better angular resolution  for SNR values above 20-30. This
is due probably to the fact that in the bright source regime the advantages of
arrays are less important, and the improved time sampling offered by
photometers becomes important. Indeed, the sampling time achieved with our
instruments (see column 5 of Table~\ref{table_list}) is 2 to 4 times slower
than what can be achieved by a fast photometer. In this respect, we hope to
optimize some aspects of the electronic and acquisition software in the near
future. In parallel, other improvements could be brought by the new generations
of detectors currently being introduced and by the increasing allocating 
time availability in larger telescopes, as discussed in Appendix~\ref{future}.

\section{Conclusions}
We have described a routine program of lunar occultation observations started
at the Spanish 1.5\,m telescope of the Calar Alto observatory, using two
relatively economical instruments for the visual and  near-IR range. A total of
40 occultations were recorded, 13 in the visual and 27 in the near--infreared,
resulting in five binary detections and 2 diameter measurements.  This being a
small sample, it is hard to infer any statistical conclusions, however it is
interesting to note that the fraction of detected binaries (5/40, of which 4/40
represent new detections) is very similar to the value of $\approx$10\%
calculated for field stars by Evans et al. (\cite{evans83}) and Richichi et al.
(\cite{richichi96}).

We have evaluated the performance of the two instruments in terms of
signal-to-noise ratio achieved for LO observations, and we conclude that their
use at the 1.5\,m telescope enables to carry out discoveries of binaries among
field stars down to the limit of about 8-9\,mag in the R and K  bands.  We have
compared these observations with the database of near-infrared lunar
occultations recorded with a fast InSb photometer at the 1.5\,m TIRGO telescope
(Richichi et al. \cite{richichi96}), and we conclude that the use of an array
detector offers significantly better sensitivity.

\begin{acknowledgements} This work was supported in part by the DGICYT
Ministerio de Ciencia y Tecnolog\'{\i}a (Spain) under grant no. AYA2001-3092.
O. Fors was supported by a fellowship from DGESIC Ministerio de Educaci\'{o}n,
Cultura i Deportes (Spain), ref. AP97~38107939. We thank the Observatorio
Astronomico Nacional for the facilities and support made available at Calar
Alto. We would like to express our gratitude to Christoph Flohr for making
available his program SCAN. We thank David Dunham and Dave Herald for
\object{SAO~78168} information supplied in private communication. This research
has made use of the \textit{USNOFS Image and Catalogue Archive} operated by the
United States Naval Observatory, Flagstaff Station ({\tt
http://www.nofs.navy.mil/data/fchpix/}). This project makes use of data
products from the \textit{Two Micron All Sky Survey}, which is a joint project
of the University of Massachusetts and the Infrared Processing and Analysis
Center/California Institute of Technology, funded by NASA and the NSF. This
research has made use of the \textit{Simbad} database, operated at CDS,
Strasbourg (France).  \end{acknowledgements}

\appendix

\section{Improvements and future possibilities}\label{future}
The technologies involved in both CCD and IR arrays manifacturing are under
continuous improvement, thus producing detectors with better performance.

For what concerns CCDs,  two important achievements have recently occurred.
Firstly, subelectron readout noise has been achieved for the first time thanks 
to low--light chip  technology (e2v Technologies \cite{e2v02}). This has been a
major step forward for low signal applications such as LO. Secondly, on-board
image memory has been recently implemented in commercial cameras (Apogee
\cite{apogee03}), permitting to record fast frame sequences without being
limited by the data transfer interface throughput (USB, Ethernet, etc).  This
is of crucial importance for LO observations in the visual, since it opens the
possibility of recording occultations at millisecond rates on  the basis of
real subframe mode, as opposed to the drift-scanning techniques which only
records the flux on one pixel.

Catering to  the needs of adaptative optics, new IR arrays with readout noise
of very few electrons are presently being introduced. In addition, faster
on-board image storage memory will allow next-generation arrays to increase the
time sampling of the occultation to 1 or 2 milliseconds, yielding an
improvement in limiting resolution.

As a result, a significant improvement in SNR and resolution is to be expected
in the near future.  In particular, it is hoped that such technological
achievements  will be transferred to a wide range of detectors, and become
available also to relatively low-budget programs such as the one here
described. In particular, access to fast, sensitive detectors at affordable
cost could be the key to promote LO observations not only at professional large
observatories, but also at smaller facilities. In turn, this could partially
overcome some of the intrinsic limitations of LO, such as the lack of repeated
observations at various wavelengths, epochs and position angles.

In parallel to the improvements expected from detector developments, also the
trend in telescope availability holds promises of increased LO performance in
the near future. While LO have the advantage of providing an  angular
resolution which is not limited by the diffraction limit of the telescope, the
technique is of course not unsensitive to benefits of observing with large
facilities.  In particular for the case of binary stars, the increase in SNR
achieved by moving to a large telescope is reflected directly in the range of
brightness ratios of possible companions that can be explored, and also extends
dramatically the number of stars that can be studied.

The extrapolation of LO observations to larger telescopes can be split in two
diameter regimes. Telescopes in the  3-4\,m class offer growing availability
due to the increasing number of 8-10\,m telescopes. The forthcoming 30-100\,m 
facilities will accentuate even more this trend. By making use of flexible time
allocation schemes, routine LO observations could be implemented with this
class of telescopes. Typically, a 3.5\,m telescope would offer a limiting
magnitude gain of about 1.5 units in K with respect to the facility used in the
current paper (Richichi \cite{richichi94}). Telescopes in the  8-10\,m class 
could be used for special opportunities, and achieve a performance of the
utmost quality. For example, Richichi (\cite{richichi03}) has investigated the
possibility to use LO at very large telescopes to perform detailed studies of
stars with exoplanets candidates. In the years 2004 to 2008), up to 14 events
could be observed from the largest observatories.

\end{document}